# Josephson Junctions Without Pairing?

Alan M. Kadin

*Abstract*—It is well established that superconductivity is based on a coherent quantum state of Cooper pairs with charge 2*e*, and this is equally true of the Josephson effect. In contrast, Kadin recently presented an alternative real-space model of the superconducting ground state, in which each electron is a localized orbital, on the scale of the coherence length ξ, coupled to a dynamic charge (or spin) density wave. A supercurrent corresponds to collective motion of a dense orthogonal array of such electrons. The present paper extends this analysis to two weakly coupled superconductors, and shows how the Josephson effect (including the ubiquitous factor of *h*/2*e*) may be obtained. Furthermore, much of the standard BCS formalism on the macroscale follows from this microstructure. However, major deviations from standard Josephson behavior are predicted for nanoscale SQUIDs much smaller than ξ.

## I. INTRODUCTION

SUPERCONDUCTIVITY is universally recognized as a unique physical phenomenon associated with quantum effects on the macroscopic scale. Key among these are quantization of magnetic flux in superconducting loops in units of $\Phi_0 = h/2e$, as well as the Josephson effect, with an oscillating supercurrent at a frequency $f_J = V/\Phi_0$. These phenomena are easily understood if a superconductor has a macroscopic wavefunction based on a superposition of identical microscopic quantum wavefunctions, each corresponding to a charge of 2*e* [1]. Furthermore, it has long been known that fundamental quantum particles are of two types: fermions (such as electrons) with half-integer spin that are forbidden to have the same wavefunction, and bosons with integer spin that are permitted to do so. A bound state of two fermions may act like a single boson. So how can a superconductor composed of electrons act like a boson condensate? Easy, just combine the electrons into bound pairs, then condense the pairs into a macroscopic quantum state. This was the motivation for the Cooper pair [2], which subsequently provided the basis for the 1957 theory of superconductivity of Bardeen, Cooper, and Schrieffer (BCS) [3]. Given the subsequent experimental verification of *h*/2*e* and other aspects of the BCS theory, the pairing basis for superconductivity has seldom been seriously questioned.

Despite this universal acceptance of pairing, there has never been a clear microscopic real-space picture of Cooper pairs in the superconducting ground state. The size of a Cooper pair is of order the superconducting coherence length ξ (~100 nm for conventional superconductors), and there may be millions of such pairs overlapping within this volume. This is inconsistent with a picture of discrete bound pairs, with the internal binding based on interactions with phonons.

In contrast, a new alternative picture of the superconducting ground state [4] consists of a dense array of localized electron orbitals that preserves long range phase coherence, without pairing. This orthogonal fermion packing (see Fig. 1) can be illustrated by a checkerboard array, where each red and black square represents an identical electron wavefunction, but with 180° (π) phase shift between nearest neighbors to preserve orthogonality. This two-phase picture may seem more appropriate for an insulator than for a superconductor, but the present paper will show not only that this permits one to construct a superconducting state, but that *h*/2*e* is reproduced on the macroscopic scale for flux quantization and the Josephson effect.

Furthermore, the localized orbitals in this two-phase microstructure can be constructed based on coherent, dynamic charge density waves (CDWs) that can be mapped onto the proven BCS formalism [5,6]. This is a remarkable result, and suggests that while the pairing interpretation of BCS provides a convenient approximation on the macroscopic scale, this orthogonal fermion packing may provide a more accurate microscopic picture of superconducting phenomena. Finally, this predicts deviations from the standard pair-based results for superconducting structures and Josephson junctions on the nanometer scale, smaller than ξ.

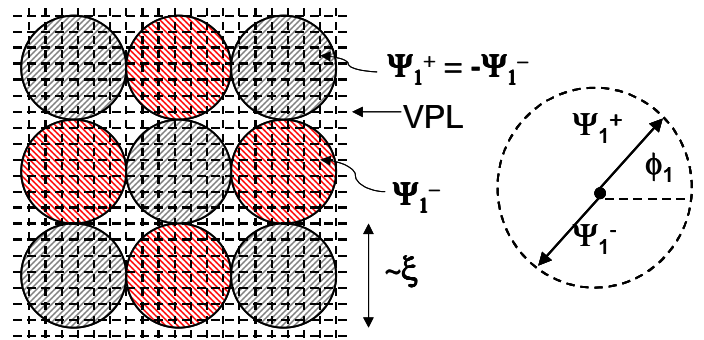

Fig. 1. Real-space picture of superconducting ground state with orthogonal packing of localized single electron orbitals (~ξ) bound to virtual-phonon lattice (VPL). Each electron is surrounded by nearest neighbors that are 180° out of phase, corresponding to a bi-phasor order parameter (right). On the macroscale, this leads to long-range phase coherence with *h*/2*e* in flux quantization and the Josephson effect.

Manuscript received 3 August 2010.
A. M. Kadin, 4 Mistflower Lane, Princeton Junction, NJ 08550 USA. Also with Hypres, Inc., 175 Clearbrook Road, Elsmford, NY 10523 USA (e-mail amkadin@alumni.princeton.edu).

## II. SUPERCONDUCTIVITY AND CHARGE DENSITY WAVES

In the standard electron band theory of crystals [7], an energy gap forms at the Brillouin zone boundary, corresponding to electron wave reflection from the periodic lattice potential, forming standing waves with charge densities that align with this potential. A solid with a Fermi level in this energy gap is an insulator at $T = 0$. The electron states just below the gap are localized in space and fixed relative to the lattice. In contrast, if the states at the Fermi level are not near a Brillouin zone boundary, the resulting Bloch waves can propagate freely through the crystal, and are delocalized, constituting metallic conduction.

However, in certain metals a new lattice distortion can be induced by the electrons near the Fermi surface, corresponding to a periodic lattice potential that reflects these electrons, and creating a new energy gap at the Fermi surface. This coupling of an electron standing wave with an induced lattice distortion constitutes a CDW [8], and is seen in certain quasi-one-dimensional (1D) conductors such as $NbSe_3$ below a critical temperature $T_c$. This CDW may be incommensurate with the crystal lattice, so in principle the CDW may move coherently relative to the fixed lattice, carrying electrical current without loss, due to the energy gap suppressing incoherent scattering. Indeed, Fröhlich proposed in 1954 [9,10] (before BCS) that such a CDW might provide the basis for superconductivity. However, it was subsequently found experimentally that all such CDW materials are in fact insulators in the limit of small voltages, due to pinning of the lattice distortion on impurities. Application of a voltage can depin the CDW distortion, leading to coherent current flow known as a sliding CDW [8]. More recently, both theory and experiment have shown that there is a close connection between CDWs and the BCS theory, despite the lack of electron pairing in CDWs. Specifically, the CDW energy gap maps onto the BCS gap equation, with the same electron-phonon coupling constant [8,11].

A CDW is based on a static 1D lattice distortion that is sometimes called a "frozen phonon". In contrast, ref. [4] considers a set of dynamic distortions in 3D that are effectively standing-wave phonons. These coherent 3D oscillations constitute a "virtual phonon lattice" (VPL), incommensurate with the crystal lattice. Unlike the static distortions of conventional CDWs, these dynamic distortions will not pin on impurities, and thus are compatible with true superconductivity. This picture does not uniquely specify the relevant phonon frequency, but the recent observations of Kohn anomalies [12] in the phonon spectra of Nb and Pb [13] suggests that this may be at the gap frequency $2\Delta/h$.

Within this dynamic CDW ground state, the electrons just below the Fermi surface have a range of different energies, but all such electron states (of both spins) are localized orbitals on the scale of the coherence length $\xi$, and all reinforce the same real-space VPL. This is directly analogous to the way that all Bloch waves in the valence band of a crystal have charge densities that are in phase with the same crystal lattice [7].

While this analysis has focused on CDWs, a closely related phenomenon is a static distortion in the lattice spin density known as a spin density wave (SDW) [14]. Static SDW transitions have been observed in several magnetic materials. A direct generalization of the dynamic CDW is to a dynamic SDW (a standing-wave magnon) that may also act as the basis for localized electrons in the superconducting ground state [4,6]. This may be relevant to high-temperature superconducting materials that are associated with magnetic excitations rather than with phonons.

## III. BI-PHASOR ORDER AND FLUX QUANTIZATION

Earlier efforts to explain superconductivity by way of CDWs [9,10] could account for zero resistance (due to a moving condensate with an energy gap), but not other effects (such as the Meissner effect and flux quantization) which require long-range phase coherence. In contrast, the pairing interpretation of BCS naturally leads to a macroscopic quantum wavefunction that derives from the superposition of identical microscopic wavefunctions. One cannot obtain this by simple superposition of electron wavefunctions, which are subject to the Pauli exclusion principle. How can one obtain long-range phase coherence from a dense array of electrons with different energies?

The key is the orthogonal two-phase packing shown in Fig. 1. In a fully occupied ground state, all electrons in the same location must be orthogonal states. If they have different energies (or spins), this orthogonality is automatically achieved. If they have the same energy and spin, they will tend to pack as densely as possible without overlapping. The only way for adjacent orbitals of the same energy to remain orthogonal is for them to be 180° out of phase, i.e., opposite signs with a node between them. In 2D, each square in a checkerboard is surrounded by four nearest neighbors of the other color sub-lattice. This extends to 3D in a cubic NaCl structure with 6 nearest neighbors, and indeed is directly analogous to the packing of ions in an ionic crystal. (Other structures for ionic crystals include the ZnS structure with 4 nearest neighbors and the CsCl structure with 8 nearest neighbors.) Because of the interactions of the two sub-lattices, the quantum phase of each sub-lattice can be maintained over macroscopic distances. Rather than a single quantum wavefunction $\Psi_2 \propto \exp(i\phi_2)$ for a pair condensate, here one has two single-electron phase factors that one might call a bi-phasor (see Fig. 1):

$$\Psi_1^+ \propto \exp(i\phi_1); \quad \Psi_1^- \propto \exp(i\phi_1+\pi) = -\exp(i\phi_1). \quad (1)$$

Strictly speaking this bi-phasor coherence applies separately to each distinct energy level (and spin); different energy levels are mutually incoherent. However, gradients in phase $\nabla\phi_1$ *are* coherent among the various energy levels, since all energy levels are tied to the same VPL and move together coherently with the same velocity $v_s$. As usual for a quantum wave of mass m and charge $-e$ in a magnetic vector potential $A$,

$$\hbar\nabla\phi_1 = mv_s - eA. \quad (2)$$

If one extracts the time-dependent phase changes due to the different energy levels, one can identify the same macroscopic bi-phasor order parameter as in Eq. (1), for all energy levels. Furthermore, since Eq. (2) for the pair condensate would have a factor of two on the right, the macroscopic $\phi_1$ and $\phi_2$ are related by $\phi_1 = \phi_2/2$. Flux quantization can be derived by considering a superconducting wire connected into a closed loop containing magnetic flux $\Phi = \int(A \cdot d\ell)$, with $v_s = 0$ in the interior of the wire. Now if one integrates Eq. (2) around the loop, one obtains a phase difference $\Delta\phi_1 = e\Phi/\hbar$. But since the bi-phasor order parameter is not single-valued but rather double-valued, $\Psi_1^+$ can connect to either $\Psi_1^+$ or $\Psi_1^-$. Then one has $\Delta\phi_1 = n\pi$, for any integer $n$, leading to the same result as for a pair wavefunction:

$$\Phi = nh/2e = n\Phi_0. \tag{3}$$

This derivation implicitly assumes that the superconducting loop is much larger than $\xi$. Implications of smaller loops will be considered in Section V.

## IV. JOSEPHSON TUNNELING IN BI-PHASOR MODEL

There are three key aspects of the Josephson effect in a classic tunnel junction that an alternative picture must reproduce. First, for $V=0$ across the junction, the dc interference effect must go as $\sin(\Delta\phi_2) = \sin(2\Delta\phi_1)$, where $\Delta\phi_1$ is the one-electron phase difference and $\Delta\phi_2$ is the two-electron phase difference. Second, for $V>0$, the supercurrent must oscillate with frequency $f_J = V/\Phi_0$. And third, the junction critical current $I_c$ must scale in the same way as the normal-state junction conductance $G_N = 1/R_N$, i.e., $I_c R_N$ must be independent of the tunnel barrier.

For reference, first consider a normal-state NIN TJ with a barrier width $d$ and a WKB decay constant $\kappa$ for a single-electron wavefunction. The tunneling amplitude from a full state on one side of the barrier to an empty state on the other goes approximately as $\Psi_1(d) \propto \exp(-\kappa d)$, so the total tunneling current has the dependence

$$I_N \propto V |\Psi_1(d)|^2 \propto V \exp(-2\kappa d). \tag{4}$$

The tunneling resistance is then

$$R_N = 1/G_N = V/I_N \propto \exp(+2\kappa d). \tag{5}$$

Before evaluating the Josephson current for the bi-phasor wavefunction, first review a simple derivation for the conventional pair Josephson current [1]. The dependence of a pair wavefunction $\Psi_2(x)$ must scale as the square of a single-electron wavefunction, so within the barrier $\Psi_2(x) \propto \exp(-2\kappa x)$, and the pair wavefunctions on the left and right sides is:

$$\Psi_2^L = C \exp[-2\kappa(x+d/2)] \exp(i\phi_2^L);$$
$$\Psi_2^R = C \exp[+2\kappa(x-d/2)] \exp(i\phi_2^R), \tag{6}$$
$$I_{s2} \propto (-e\hbar/2m) \operatorname{Im}[\Psi_2^{T*} \nabla \Psi_2^T] \propto \exp(-2\kappa d) \sin(\phi_2^L - \phi_2^R)$$
$$= I_{c2} \sin(\Delta\phi_2). \tag{7}$$

So $I_{c2} \propto \exp(-2\kappa d)$, leading to $I_{c2} R_N$ independent of $\kappa$ and $d$. Note that the Josephson current does <u>not</u> correspond to tunneling of pairs from a full state on one side to an empty state on the other. That would scale as $|\Psi_2(d)|^2 \propto \exp(-4\kappa d)$, which would have the wrong scaling with $d$.

This conventional derivation does not require that the pair wavefunctions on the two sides be at the same voltage. For a voltage $V$ across the junction, the quantum phase difference evolves as $\hbar d(\Delta\phi_2)/dt = -\Delta E_2 = 2eV$. This leads in the usual way to the ac Josephson effect with a current oscillation

$$I_{s2} = I_{c2} \sin(2\pi f_J t), \tag{8}$$

at the Josephson frequency $f_J = 2eV/h$. In addition to the electromagnetic energy difference $2eV$, there is also a Josephson energy $E_{J2}$ across the barrier, which can be derived by integrating the electrical power coupled into the system:

$$E_{J2} = \int dt\, I_{s2} V = -(\hbar I_{c2}/2e) \cos(\Delta\phi_2). \tag{9}$$

Of course, for finite voltages, there is also incoherent tunneling current due to quasiparticle states. i.e., hole states below the energy gap and electron states above. These tunnel from full to empty states, so the total quasiparticle current scales as $I_{qp} \propto \exp(-2\kappa d)$, the same as for NIN tunneling.

Now consider a similar derivation for the single-electron bi-phasor. For each energy level in the ground state, one obtains an interference current $I_{s1} = I_{c1} \sin(\Delta\phi_1)$ where $I_{c1} \propto \exp(-\kappa d)$, both of which would seem in conflict with experiments. However, because of the two-component nature of the waves on each side, both components on one side may interfere with either component on the other. That gives rise to two possible signs for the supercurrent, e.g., $I_{s1}^{++} = I_{c1} \sin(\Delta\phi_1)$ and $I_{s1}^{+-} = -I_{c1} \sin(\Delta\phi_1) = -I_{s1}^{++}$. Which sign will apply for a particular energy level depends on the configuration of orbitals on either side adjacent to the tunnel barrier. The total supercurrent $I_{s1}^T$ should be a linear combination of these two terms,

$$I_{s1}^T = f I_{s1}^{++} + (1-f) I_{s1}^{+-} = (2f-1) I_{s1}^{++}, \tag{10}$$

where $f$ is a fraction corresponding to ++ (or --) alignment on the two sides. But in the limit of weak coupling across the barrier, the patterns on the two sides should be uncorrelated.

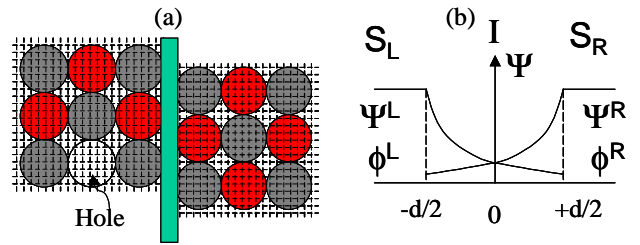

Fig. 2. Two-phase quantum interference in Josephson junction. (a) Real-space picture near tunnel barrier, showing uncorrelated packing patterns on two sides of barrier. (b) Functional dependence of wavefunctions in barrier.

For all energy levels, one then expects on average that $\langle f \rangle = \frac{1}{2}$, so that $\langle I_{s1}^T \rangle = 0$, completely cancelling this dc Josephson current on the macroscopic scale. That negates the improper scaling above, but now does not account for the dc Josephson effect.

But consider a higher-order effect, whereby the weak Josephson interaction energy across the tunnel barrier $E_{J1} = -(\hbar I_{c1}/e) \cos(\Delta\phi_1)$ (similar to Eq. 9) will tend to align the patterns on the two sides to decrease the total energy. Assume further that this is a small effect in the linear response regime, and that the average alignment fraction $\langle f - \frac{1}{2} \rangle \propto E_{J1}$. Then the relation for the bi-phasor supercurrent becomes

$$I_{s1}^{(2)} \propto E_{J1}^{++} I_{s1}^{++} \propto I_{c1}^2 \sin(\Delta\phi_1) \cos(\Delta\phi_1)$$
$$\propto \exp(-2\kappa d) \sin(2\Delta\phi_1). \qquad (11)$$

Remarkably, this 2nd order single-electron Josephson effect yields exactly the same dependences as for the 1st order pair Josephson effect, at least on a macroscopic scale:

$$I_{s1}^{(2)} = I_{c1}^{(2)} \sin(2\Delta\phi_1) = I_{c1}^{(2)} \sin(\Delta\phi_2), \qquad (12)$$

and $I_{c1}^{(2)} R_N$ is independent of $\kappa$ and $d$. The case of a nanoscale Josephson junction will be different, and is discussed further below.

The non-zero-voltage behavior of the Josephson junction should also reproduce that of the conventional theory. First, the ac Josephson effect is unchanged, given the phase dependence in Eq. (12). For a voltage $V$, $d(\Delta\phi_1)/dt = eV$, so that $I_{s1}^{(2)} = I_{c1}^{(2)} \sin(2eVt/\hbar)$ as expected. Second, holes and electrons in this picture are also localized states like the ground state orbitals (see Fig. 2a), but tunnel incoherently across the barrier with the same tunneling amplitude as do single electrons in an NIN junction. So one would expect the I-V curves to follow the conventional results, as described for SIS tunnel junctions.

The above analysis has focused on classic SIS Josephson junctions, while practical Josephson devices are often based on conductive barriers (SNS) or superconducting micro-constrictions. But the derivation does not strictly depend on the insulating barrier, only on weak coupling of two strong superconductors, each with a bi-phasor order parameter and corresponding orbital pattern. Therefore, one would generally expect the standard results to carry over to these junction types, as well.

## V. NANOSCALE JUNCTIONS AND SQUIDs

It was shown above that the bi-phasor microscopic model maps onto the conventional Josephson effect for macroscopic junctions, where one can average over the microstructure associated with the orbital packing on the scale of the coherence length $\xi$. But this mapping must break down in the nanoscale limit, where the interference is really between single orbitals, and the bi-phasor nature is not applicable. For example, consider the superconducting nanostructured loops illustrated in Fig. 3. In Fig. 3a, without a Josephson junction, each orbital is interfering with itself, which should yield flux periodicity $h/e$ associated with a single-electron orbital [4]. A recent theoretical analysis ostensibly based on conventional theory has come to the same conclusion [15]. There have also recently been magnetic measurements on very small superconducting loops and arrays [16], but these may not be quite small enough to test this prediction. Fig. 3b shows a similar loop with a small junction, and represents an rf SQUID. This corresponds again to self-interference of single orbitals with $h/e$ flux periodicity. Furthermore, $I_c$ should also be 1st order, larger than the standard value based on $R_N$ of the junction. The 2-junction dc SQUID in Fig. 3c corresponds to local interference between two different orbitals, but again the loop is too small for averaging over orbital patterns. A similar 1st order response would be expected. There have recently been nano-SQUIDs fabricated from superconducting nanobridges [17-20]. There have been as yet no experimental observations of $h/e$ periodicity in these nano-SQUIDs, but these loops may not yet be in the limit $\ll \xi$.

The SQUID loops in Fig. 3 should be distinguished from measurement of a nano-scale junction $\ll \xi$ embedded in a macroscopic circuit, where the phase $\Delta\phi_1$ is a macroscopic variable that reflects the bi-phasor order parameter. This would lead to cancellation of the 1st order Josephson effect associated with $h/e$. However, it is difficult to see how one obtains the induced pattern alignment in such a nano-junction, to obtain the 2nd order effect in Eq. (11) that reproduces the standard result.

## VI. CONCLUSION

The present paper has applied a novel real-space microscopic model of the superconducting ground state to the classic Josephson effect in tunnel junctions, and has shown that most of the conventional results are reproduced, even though this is a single-electron band theory without Cooper pairs. This is a consequence of spatial averaging of a bi-phasor order parameter, which follows from dense orthogonal fermion packing of localized electrons in the ground state. But the real test of such an alternative model is in cases that predict different results. The should occur on scales much smaller than $\xi$, for which the spatial averaging cannot be applied. In particular, SQUID loops on this nanoscale should exhibit a flux periodicity $h/e$ corresponding to a single-electron wavefunction, together with an enhanced critical current. Measurements to date on nano-SQUIDs may not yet be quite in this regime, but future experimental and theoretical research may help to address this important question about the microscopic foundations of the Josephson effect.

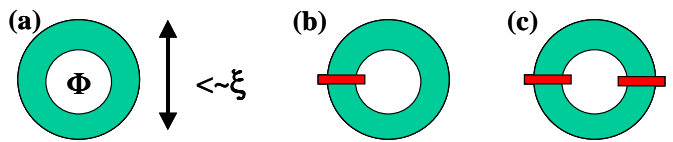

Fig. 3. Superconducting nanostructures predicted to exhibit $h/e$ flux periodicity. (a) Superconducting loop. (b) Single-junction RF SQUID. (c) Two-junction DC SQUID.